\documentclass{moriond}

\usepackage{amsmath}
\usepackage{amssymb}

\def\be{\begin{equation}}
\def\ee{\end{equation}}
\def\bea{\begin{eqnarray}}
\def\eea{\end{eqnarray}}

\newcommand{\nn}{\nonumber}

\newcommand{\Photo}{}

\begin{document}
\vspace*{4cm}
\title{EFFECTIVE FIELD THEORY OF POST-NEWTONIAN GRAVITY INCLUDING SPINS}

\author{MICHELE LEVI}

\address{Institut de Physique Th\'eorique, Universit\'e Paris-Saclay, CEA, CNRS,\\ 
91191 Gif-sur-Yvette, France}

\maketitle\abstracts{
We present in detail an Effective Field Theory (EFT) formulation for the essential case of 
spinning objects as the components of inspiralling compact binaries. We review its 
implementation, carried out in a series of works in recent years, which leveled the 
high post-Newtonian (PN) accuracy in the spinning sector to that, recently attained in the 
non-spinning sector. We note a public package, ``EFTofPNG'', that we recently 
created for high precision computation in the EFT of PN Gravity, which covers all sectors, 
and includes an observables pipeline.
}

\section{Introduction}

A new era of high precision Gravity has been launched with the recent direct detection of 
gravitational waves (GWs) from compact binary coalescence 
\cite {Abbott:2016blz,Abbott:2016nmj}. The influx of improved GW data is expected to 
increase, and more accurate GWs templates will be required to optimally analyze the signal. 
The continuous GW signal comprises various kinds of physics, corresponding to the different 
phases in the evolution of the radiating binary. The initial inspiral phase, where the 
components of the binary orbit with a non-relativistic velocity, can only be treated 
analytically with the post-Newtonian (PN) theory of General Relativity 
\cite{TheLIGOScientific:2016src,Blanchet:2013haa}. Indeed, in recent years there has been a 
remarkable progress in high order PN theory, in particular also involving the essential 
\textit{spins} of the components of the binary. In what follows we detail the formal 
progress, which has been obtained within an extension of the Effective Field Theory (EFT) 
for the binary inspiral to spinning objects \cite{Levi:2010zu,Levi:2015msa}. We also review 
its specific applications, which we carried out.

\section{Effective Field Theory of Post-Newtonian Gravity including Spinning Objects}

Effective Field Theories (EFTs) and their setup are universal. Once a hierarchy of scales is 
identified in the physical problem, be it classical or quantum, the robust EFT framework can 
then be applied. For the compact binary coalescence in the inspiral phase of its  
evolution, this observation was made by Goldberger 
\textit{et al}.~\cite{Goldberger:2004jt,Goldberger:2007hy}. Indeed, there are three distinct 
characteristic length scales in the problem: $r_s$, $r$, and $\lambda$, corresponding to the 
scales of the internal structure of the single compact component of the binary, the orbital 
separation of the binary, and the wavelength of radiation emitted from the binary, 
respectively. They are widely separated by powers of $v\ll 1$, the typical 
non-relativistic orbital velocity at the inspiral phase, as we have that 
$r_s\sim r\,v^2 \sim \lambda \,v^3$. Goldberger 
\textit{et al}.~\cite{Goldberger:2004jt,Goldberger:2007hy} put forward a program to tackle 
the binary inspiral problem with a tower of EFTs, corresponding to each of the scales. We 
note that the only scale in the full theory is the UV scale $m$, the mass of the isolated 
compact object, where $r_s\sim m$. For the general case that we tackle in our work, where 
the compact object is \textit{spinning}, and hence also characterized by its spin length 
\cite{Levi:2015msa}, $S^2$, we also have that $S\lesssim m^2$.

The construction of an EFT follows one of two general procedures, top-down and bottom-up. 
These approaches differ on how the effective action, which formally represents the EFT, is 
obtained. The bottom-up approach constructs the effective action from scratch as an infinite 
sum of operators, constrained by symmetry considerations. The top-down approach obtains the 
effective action by explicitly eliminating degrees of freedom (DOFs) from the action of the 
high energy (small scale) theory. We apply, in fact, both of these approaches in our 
construction of the EFTs. First, we remove the scale $r_s$, using a bottom-up approach, by 
eliminating the strong field modes $g_{\mu\nu}^s$, where the gravitational field is 
decomposed into $g_{\mu\nu}\equiv g_{\mu\nu}^s + \bar{g}_{\mu\nu}$, and $\bar{g}_{\mu\nu}$ 
represents the field modes in scales above $r_s$. The effective action for an isolated 
compact object can then be generically written as:
\be\label{ssing}
S_{\text{eff}}\left[\bar{g}_{\mu\nu},y^\mu,e^{\mu}_{A}\right]=
-\frac{1}{16\pi G} \int d^4x \sqrt{\bar{g}} R\left[\bar{g}_{\mu\nu}\right] 
+ \underbrace{\sum_{i}C_i\int d\sigma O_i(\sigma)}_{S_{pp}\equiv
\text{point particle action}},
\ee 
where we introduce a point particle action, $S_{\text{pp}}$, with an infinite tower of 
worldline operators, $O_i(\sigma)$. These should contain the DOFs relevant at this scale, 
and satisfy the symmetries of the theory. Hence, it is crucial to accurately identify the 
DOFs and symmetries for the general case of spinning gravitating objects as we further 
elaborate below \cite{Levi:2015msa}. Here, $y^\mu$, and $e^{\mu}_{A}$, are the particle 
worldline coordinate, and tetrad DOFs, respectively. All of the UV physics is encapsulated 
in the Wilson coefficients, $C_i(r_s)$, in the point particle action, $S_{\text{pp}}$. 

For the next EFT in the tower, where the orbital scale $r$ is removed, we use the top-down 
approach. First, we decompose the field into 
$\bar{g}_{\mu\nu}\equiv\eta_{\mu\nu}+H_{\mu\nu}+\widetilde{h}_{\mu\nu}$, where $H_{\mu\nu}$ 
represents the field modes at the orbital scale, and $\widetilde{h}_{\mu\nu}$, the radiation 
modes. These different field modes scale with a definite power of the small PN parameter 
$v$: 
$\partial_t H_{\mu\nu}\sim \frac{v}{r} H_{\mu\nu} \,$, 
$\partial_i H_{\mu\nu}\sim \frac{1}{r} H_{\mu\nu} \,$,
$\partial_\rho\widetilde{h}_{\mu\nu}\sim \frac{v}{r} \widetilde{h}_{\mu\nu}$.
Then, we use the effective action of two compact objects, with a copy of $S_{\text{pp}}$ for 
each object, to integrate out the field modes $H_{\mu\nu}$ by computing the following 
functional integral:
\be \label{scomp}
e^{iS_{\text{eff(composite)}}\left[\widetilde{h}_{\mu\nu}, y^\mu, e^{\mu}_A\right]} 
\equiv \int {\cal{D}}H_{\mu\nu}~e^{iS_{\text{eff}}\left[\bar{g}_{\mu\nu}, y^\mu_{1}, 
y^\mu_{2}, e_{(1)}{}^{\mu}_{A}, e_{(2)}{}^{\mu}_{A}\right]}, 
\ee
taking only the tree level for the classical limit. This defines the effective action of the 
composite object with $y^\mu$, and $e^{\mu}_A$, now being its worldline coordinate, and 
tetrad, respectively. In general, we should proceed to integrate out the radiation modes, 
$\widetilde{h}_{\mu\nu}$, to get the final EFT. However, this is not required in the 
conservative sector, where no radiation modes are present. Therefore, we stress that the 
final effective action at this stage should consist of no remaining field modes at the 
orbital scale \cite{Levi:2008nh,Levi:2014sba,Levi:2015msa}. 

As we noted, in order to construct the point particle action in Eq.~\ref{ssing}, it is 
crucial to accurately identify the DOFs and symmetries of the theory \cite{Levi:2015msa}. As 
for the DOFs for a spinning object, of which we have three kinds, it is important to note 
the following: 1.~As for the gravitational field we should consider the tetrad field, 
$\eta^{ab}\tilde{e}_a{}^\mu(x)\tilde{e}_b{}^\nu(x)=g^{\mu\nu}(x)$, rather than just 
$g_{\mu\nu}(x)$; 2.~As for the particle worldline coordinate, $y^\mu(\sigma)$, the particle 
worldline position does not in general coincide with the object's ``center''; 3.~As for the 
particle worldline rotating DOFs, from the worldline tetrad,  
$\eta^{AB}e_A{}^\mu(\sigma)e_B{}^\nu(\sigma)=g^{\mu\nu}$, we define the angular velocity, 
$\Omega^{\mu\nu}(\sigma)\equiv e^\mu_A\frac{De^{A\nu}}{D\sigma}$, and add its conjugate, the 
worldline spin, $S_{\mu\nu}(\sigma)\equiv-2\frac{\partial L}{\partial\Omega^{\mu\nu}}$, as 
further DOFs. Later, we switch to the worldline Lorentz matrices, 
$\eta^{AB}\Lambda_A{}^a(\sigma)\Lambda_B{}^b(\sigma)=\eta^{ab}$, and the conjugate local 
spin, $S_{ab}(\sigma)$. 

Regarding the symmetries of the theory, it is crucial to note the additional symmetries that 
play a role for the spinning case, i.e.~beyond general covariance, and worldline 
reparametrization invariance. These additional symmetries are: 1.~Parity invariance; 
2.~Internal Lorentz invariance of the local frame field; 3.~$SO(3)$ invariance of the worldline spatial triad; 4.~Spin gauge invariance, that is invariance under the choice of completion of the worldline spatial triad through a timelike vector. This is a gauge of the rotational variables, i.e.~of both the worldline tetrad, and spin. In addition to these symmetries, we assume that the isolated object has no intrinsic permanent multipole moments beyond the mass monopole, and the spin dipole.
 
For a spinning object we can write the point particle action in Eq.~\ref{ssing} in
the form \cite{Hanson:1974qy,Bailey:1975fe,Levi:2015msa}:
\begin{align}\label{spp}
S_{\text{pp}} = \int d \sigma \left[ -m \sqrt{u^2}
								- \frac{1}{2} S_{\mu\nu}\Omega^{\mu\nu}
                + L_{\text{SI}}\left[u^{\mu}, S_{\mu\nu}, \bar{g}_{\mu\nu}
                \left(y^\mu\right)\right]\right],
\end{align} 
where $u^\mu\equiv dy^\mu/d\sigma$, and $L_{\text{SI}}$ denotes
the Lagrangian part, which is nonminimal coupling, and as we assume, contains only 
spin-induced higher multipoles. While the minimal coupling here is fixed only from 
covariance and reparametrization invariance, it can still be worked out to further 
incorporate the other symmetries related with the worldline tetrad. As for nonminimal 
coupling, parity invariance also plays a role in constraining it. Indeed, in 
\cite{Levi:2015msa} we work out the point particle action of a spinning object, required for 
the first EFT of a single spinning particle. 

In the spirit of Stueckelberg, we first want to introduce the gauge freedom of the 
rotational variables into the effective action. We do this by applying a 4-dimensional 
covariant boost-like transformation on the worldline tetrad \cite{Levi:2015msa}. This 
introduces new gauge DOFs, $w_{\mu}$, as the timelike vector of the tetrad, 
$\hat{e}_{[0]\mu} = w_{\mu}$, and also leads to a generic gauge condition for the spin:
\begin{equation}\label{gensgauge}
\hat{S}^{\mu\nu} \left( p_{\nu} + \sqrt{p^2} \hat{e}_{[0]\nu} \right) = 0.
\end{equation}
All in all, these constitute the $3+3$ necessary gauge conditions for the redundant 
unphysical DOFs, that are contained in the 4-dimensional antisymmetric angular velocity, and 
spin tensors. The minimal coupling term then yields:
\begin{align}\label{mcTrans}
\frac{1}{2} S_{\mu\nu} \Omega^{\mu\nu} &= 
  \frac{1}{2} \hat{S}_{\mu\nu} \hat{\Omega}^{\mu\nu}
	+ \frac{\hat{S}^{\mu\nu} p_{\nu}}{p^2} \frac{D p_{\mu}}{D \sigma},
\end{align}
where an extra term appears in the action. This extra term, originating from minimal 
coupling, is of course not preceded by any Wilson coefficient, although it contributes to 
finite size effects. As for the spin that appears in spin-induced nonminimal couplings in the action, it is transformed to the generic spin variable as: 
\begin{equation}\label{sTrans}
S_{\mu\nu} = \hat{S}_{\mu\nu} - \frac{\hat{S}_{\mu\rho} p^{\rho} p_{\nu}}{p^2}
	+ \frac{\hat{S}_{\nu\rho} p^{\rho} p_{\mu}}{p^2}.
\end{equation} 

As we noted, the nonminimal coupling action with spin should also be constrained, using the 
full set of symmetries of the theory, that we detailed above. Based on these symmetries, and 
further considerations and properties of the problem, the leading order (LO) nonminimal 
couplings are indeed fixed to all orders in spin \cite{Levi:2015msa}: 
\begin{align} \label{sppsinmc}
L_{\text{SI}}=&\sum_{n=1}^{\infty} \frac{\left(-1\right)^n}{\left(2n\right)!}
\frac{C_{ES^{2n}}}{m^{2n-1}} D_{\mu_{2n}}\cdots D_{\mu_3}
\frac{E_{\mu_1\mu_2}}{\sqrt{u^2}} S^{\mu_1}S^{\mu_2}\cdots 
S^{\mu_{2n-1}}S^{\mu_{2n}}\nn\\
&+\sum_{n=1}^{\infty} \frac{\left(-1\right)^n}{\left(2n+1\right)!}
\frac{C_{BS^{2n+1}}}{m^{2n}} 
D_{\mu_{2n+1}}\cdots D_{\mu_3}\frac{B_{\mu_1\mu_2}}{\sqrt{u^2}} 
S^{\mu_1}S^{\mu_2}\cdots 
S^{\mu_{2n-1}}S^{\mu_{2n}}S^{\mu_{2n+1}},
\end{align}
where new spin-induced Wilson coefficients precede each of the nonminimal coupling terms. 
These operators are composed from either the electric, or magnetic curvature tensors, 
$E_{\mu\nu}$, or $B_{\mu\nu}$, respectively, together with the spin vector $S^{\mu}$. Of the 
above operators, the quadrupole, octupole, and hexadecapole couplings should notably be 
taken into account up to the fourth PN (4PN) order 
\cite{Levi:2014gsa,Levi:2015msa,Levi:2015ixa}. These couplings explicitly read:
\begin{align} \label{4pns2s3s4}
L_{ES^2} \equiv& -\frac{C_{ES^2}}{2m} \frac{E_{\mu\nu}}{\sqrt{u^2}} 
S^{\mu} S^{\nu},\\
L_{BS^3}\equiv&-\frac{C_{BS^3}}{6m^2}\frac{D_\lambda B_{\mu\nu}}
{\sqrt{u^2}}S^{\mu} S^{\nu} S^{\lambda},\\
L_{ES^4}\equiv&\frac{C_{ES^4}}{24m^3} \frac{D_\lambda D_\kappa 
E_{\mu\nu}}{\sqrt{u^2}} S^{\mu} S^{\nu} S^{\lambda} S^{\kappa}.
\end{align}

We recall that for the second EFT we need to remove the field modes at the orbital scale. To 
this end the field DOFs should be disentangled from the particle DOFs. This can only be 
attained if the gauge of the rotational variables is fixed in the action, as was put forward 
in \cite{Levi:2008nh}. We stress that as we work in an action approach the gauge of the 
rotational variables can be directly inserted at any stage. But first we need to switch to 
new rotational variables: the locally flat angular velocity with worldline Lorentz matrices, 
$\hat{\Omega}^{ab}_{\text{flat}} 
= \hat{\Lambda}^{Aa} \frac{d \hat{\Lambda}_A{}^b}{d \sigma}$, and the local spin,
$\hat{S}_{ab}=\tilde{e}^{\mu}_{a}\tilde{e}^{\nu}_{b}\hat{S}_{\mu\nu}$. Then, using the Ricci 
rotation coefficients with the tetrad field, defined by
$\omega_{\mu}{}^{ab} \equiv \tilde{e}^b{}_{\nu} D_{\mu}\tilde{e}^{a\nu}$, 
we can rewrite the minimal coupling term in Eq.~\ref{mcTrans} in the form 
\cite{Levi:2010zu,Levi:2015msa}:
\begin{align} \label{ssplitfield}
\frac{1}{2} \hat{S}_{\mu\nu} \hat{\Omega}^{\mu\nu}&= 
    \frac{1}{2} \hat{S}_{ab} \hat{\Omega}^{ab}_{\text{flat}}
         + \frac{1}{2} \hat{S}_{ab} \omega_{\mu}{}^{ab} u^{\mu}.      
\end{align}
Now, before we integrate out the field modes at the orbital scale, we need to fix all gauges 
in the action, which also eliminates all unphysical DOFs from the action. To begin with, we 
apply the beneficial non-relativistic space+time Kaluza-Klein decomposition on the 
gravitational field, i.e.~we switch to the non-relativistic gravitational (NRG) fields 
\cite{Kol:2007bc,Kol:2010ze}. The tetrad field gauge is then accordingly fixed to 
Schwinger's time gauge with the NRG parametrization, and finally we fix the gauge of the 
rotational variables to a gauge, that is dubbed the ``canonical'' gauge.

\section{Summary of formal results and applications}

In conclusion, we have provided an EFT formulation for the essential case of spinning 
objects as the components of inspiralling compact binaries, which constitutes a challenging 
extension of the EFT of PN Gravity \cite{Levi:2015msa}. In particular, we have also provided 
spin-induced nonminimal couplings to all orders in spin for a gravitating spinning particle. 
Moreover, the equations of motion (EOMs), and Hamiltonians are simply derived from the 
resulting effective action of the composite object. The (physical) EOMs of both the 
position, and spin, take a simple form, and are obtained directly via a proper variation of 
the action. Notably, for the precession equations it should be stressed, that one makes an 
independent variation with respect to the spin, and to its conjugates, the Lorentz matrices 
\cite{Levi:2014sba,Levi:2015msa}. In addition, the useful Hamiltonians are obtained in the 
standard manner similarly to the non-spinning case. 

As for the implementation, we have completed in a series of works all of the interaction 
potentials, as well as their derivatives and observables, in the spinning sectors up to the 
4PN order, on par with the PN accuracy recently attained in the generally simpler point mass 
sector. These potentials include: the linear-in-spin next-to-next-to-leading order (NNLO) 
spin-orbit at 3.5PN order \cite{Levi:2015uxa}, and NNLO spin1-spin2 at 4PN order 
\cite{Levi:2011eq,Levi:2014sba}, the NNLO spin-squared at 4PN order 
\cite{Levi:2015ixa,Levi:2016ofk}, and the LO cubic, and quartic in spin at 3.5PN, and 4PN 
orders, respectively \cite{Levi:2014gsa}.

Moreover, we have recently created a new public package, ``EFTofPNG'', for high precision 
computation in the EFT of PN Gravity (PNG), including spins \cite{Levi:2017kzq}. The 
``EFTofPNG'' package version 1.0 covers the point mass sector, and all the spin sectors, up 
to the 4PN order, and two-loop level. It is released as a public repository in GitHub, and 
can be found in the URL: ``https://github.com/miche-levi/pncbc-eftofpng''. The ``EFTofPNG'' 
package is self-contained, modular, and designed to be accessible to the classical Gravity 
community. Its final unit provides the full computation of derivatives of interest, and 
gauge invariant observables, and serves as a pipeline chain for the modeling of GW templates 
for the detectors.

\section*{Acknowledgments}

ML acknowledges the prolific collaboration with Jan Steinhoff on this line of research.
ML would like to thank John Joseph Carrasco for his constant support.
ML is grateful for the generous grant from the Moriond conference.
The work of ML is supported by the European Research Council under the European Union's 
Horizon 2020 Framework Programme FP8/2014-2020 grant no.~639729, preQFT project.

\section*{References}

\bibliography{gwbibtex}

\begin{thebibliography}{10}

\bibitem{Abbott:2016blz}
B.~ P Abbott et~al.
\newblock {Observation of Gravitational Waves from a Binary Black Hole Merger}.
\newblock {\em Phys. Rev. Lett.}, 116:061102, 2016.

\bibitem{Abbott:2016nmj}
B.~P. Abbott et~al.
\newblock {GW151226: Observation of Gravitational Waves from a 22-Solar-Mass
  Binary Black Hole Coalescence}.
\newblock {\em Phys. Rev. Lett.}, 116:241103, 2016.

\bibitem{TheLIGOScientific:2016src}
B.~P. Abbott et~al.
\newblock {Tests of general relativity with GW150914}.
\newblock {\em Phys. Rev. Lett.}, 116:221101, 2016.

\bibitem{Blanchet:2013haa}
Luc Blanchet.
\newblock {Gravitational Radiation from Post-Newtonian Sources and Inspiralling
  Compact Binaries}.
\newblock {\em Living Rev.Rel.}, 17:2, 2014.

\bibitem{Levi:2010zu}
Michele Levi.
\newblock {Next to Leading Order gravitational Spin-Orbit coupling in an
  Effective Field Theory approach}.
\newblock {\em Phys.Rev.}, D82:104004, 2010.

\bibitem{Levi:2015msa}
Michele Levi and Jan Steinhoff.
\newblock {Spinning gravitating objects in the effective field theory in the
  post-Newtonian scheme}.
\newblock {\em JHEP}, 09:219, 2015.

\bibitem{Goldberger:2004jt}
Walter~D. Goldberger and Ira~Z. Rothstein.
\newblock {An Effective field theory of gravity for extended objects}.
\newblock {\em Phys.Rev.}, D73:104029, 2006.

\bibitem{Goldberger:2007hy}
Walter~D. Goldberger.
\newblock {Les Houches lectures on effective field theories and gravitational
  radiation}.
\newblock arXiv:hep--ph/0701129, [hep-ph].

\bibitem{Levi:2008nh}
Michele Levi.
\newblock {Next to Leading Order gravitational Spin1-Spin2 coupling with
  Kaluza-Klein reduction}.
\newblock {\em Phys.Rev.}, D82:064029, 2010.

\bibitem{Levi:2014sba}
Michele Levi and Jan Steinhoff.
\newblock {Equivalence of ADM Hamiltonian and Effective Field Theory approaches
  at next-to-next-to-leading order spin1-spin2 coupling of binary inspirals}.
\newblock {\em JCAP}, 1412:003, 2014.

\bibitem{Hanson:1974qy}
Andrew~J. Hanson and T.~Regge.
\newblock {The Relativistic Spherical Top}.
\newblock {\em Annals Phys.}, 87:498, 1974.

\bibitem{Bailey:1975fe}
I.~Bailey and W.~Israel.
\newblock {Lagrangian Dynamics of Spinning Particles and Polarized Media in
  General Relativity}.
\newblock {\em Commun.Math.Phys.}, 42:65--82, 1975.

\bibitem{Levi:2014gsa}
Michele Levi and Jan Steinhoff.
\newblock {Leading order finite size effects with spins for inspiralling
  compact binaries}.
\newblock {\em JHEP}, 06:059, 2015.

\bibitem{Levi:2015ixa}
Michele Levi and Jan Steinhoff.
\newblock {Next-to-next-to-leading order gravitational spin-squared potential
  via the effective field theory for spinning objects in the post-Newtonian
  scheme}.
\newblock {\em JCAP}, 1601:008, 2016.

\bibitem{Kol:2007bc}
Barak Kol and Michael Smolkin.
\newblock {Non-Relativistic Gravitation: From Newton to Einstein and Back}.
\newblock {\em Class.Quant.Grav.}, 25:145011, 2008.

\bibitem{Kol:2010ze}
Barak Kol, Michele Levi, and Michael Smolkin.
\newblock {Comparing space+time decompositions in the post-Newtonian limit}.
\newblock {\em Class.Quant.Grav.}, 28:145021, 2011.

\bibitem{Levi:2015uxa}
Michele Levi and Jan Steinhoff.
\newblock {Next-to-next-to-leading order gravitational spin-orbit coupling via
  the effective field theory for spinning objects in the post-Newtonian
  scheme}.
\newblock {\em JCAP}, 1601:011, 2016.

\bibitem{Levi:2011eq}
Michele Levi.
\newblock {Binary dynamics from spin1-spin2 coupling at fourth post-Newtonian
  order}.
\newblock {\em Phys.Rev.}, D85:064043, 2012.

\bibitem{Levi:2016ofk}
Michele Levi and Jan Steinhoff.
\newblock {Complete conservative dynamics for inspiralling compact binaries
  with spins at fourth post-Newtonian order}.
\newblock arXiv:1607.04252, [gr-qc].

\bibitem{Levi:2017kzq}
Michele Levi and Jan Steinhoff.
\newblock {EFTofPNG: A package for high precision computation with the
  Effective Field Theory of Post-Newtonian Gravity}.
\newblock arXiv:1705.06309, [gr-qc].

\end{thebibliography}

\end{document}